\documentclass[%
 reprint,
superscriptaddress,
 amsmath,amssymb,
 aps,
]{revtex4-1}
\DeclareMathOperator{\Tr}{Tr}
\DeclareMathOperator{\exponent}{e}

\usepackage{graphicx}
\graphicspath{{figures/}}  
\usepackage{dcolumn}
\usepackage{bbm}
\usepackage{braket}
\usepackage[usenames,dvipsnames,svgnames,table]{xcolor}

\begin{document}


\title{Witnessing irreducible dimension}

\author{Wan Cong}
\affiliation{Centre for Quantum Technologies, National University of Singapore, Singapore}
\author{Yu Cai}
\affiliation{Centre for Quantum Technologies, National University of Singapore, Singapore}
\author{Jean-Daniel Bancal}
\affiliation{Quantum Optics Theory Group, University of Basel, Switzerland}
\author{Valerio Scarani}
\affiliation{Centre for Quantum Technologies, National University of Singapore, Singapore}
\affiliation{Department of Physics, National University of Singapore, Singapore}

\date{\today}

\begin{abstract}

The Hilbert space dimension of a quantum system is the most basic quantifier of its information content. Lower bounds on the dimension can be certified in a device-independent way, based only on observed statistics. We highlight that some such ``dimension witnesses'' capture only the presence of systems of some dimension, which in a sense is trivial, not the capacity of performing information processing on them, which is the point of experimental efforts to control high-dimensional systems. In order to capture this aspect, we introduce the notion of irreducible dimension of a quantum behaviour. This dimension can be certified, and we provide a witness for irreducible dimension four.



\end{abstract}

\maketitle

\section{Introduction}

The Hilbert space dimension of a quantum system limits the amount of information that can be stored in it. The study of the power of fixed-dimensional systems is still topical today~\cite{deVincente16,Frenkel2015,Navascues15a}, and several experimental groups are implementing high-dimensional encoding and decoding of information \cite{dada,malik,gauthier}. Thus, for the purposes of quantum information processing, a proper certification of dimension should capture the users' capacity of exploiting that dimensionality, not just the dimension that ``is there'' --- after all, the simplest particle or a single mode of any field are already infinite-dimensional. To put it with another example: two qubits are a ququart, but merely using a source of qubits twice does not guarantee the ability of processing the information of a ququart.

The last decade has seen the rise of device-independent certification: some important properties of quantum devices can be assessed by looking only at the observed input-output statistics. A lower bound on the Hilbert space dimension can be certified in this way. Such \textit{device-independent dimension witnesses (DIDW)} exist both as prepare-and-measure schemes \cite{preparemeasure, pm2} and as Bell-type schemes \cite{brunner,PhysRevA.77.042105,tamas,Navascues15a}. But which notion of dimension do they capture?

In this paper, we first show that some existing DIDW unfortunately capture only the dimension that ``is there''. As such, they can certify high dimension while only sequential procedures are being implemented, like using a source of qubits several times and implementing classical feed-forward. Having brought this issue to the fore, we define the \textit{dimension irreducible under sequential operations}, or simply \emph{irreducible dimension}, that can be inferred from the available observations. Finally we introduce a witness of irreducible dimension four, that can be violated by a pair of ququarts and suitable measurements. This shows that one can obtain device-independent bounds for a notion of dimension more attuned to the needs of quantum information processing.

\section{Sequential violation of dimension witnesses}
\label{section:cglmpdw}

We focus on bipartite scenarios involving two non-communicating parties, Alice and Bob. Alice's possible measurements are labeled by $x\in\mathcal{X}$, and her outcomes $a\in\mathcal{A}$. Bob's measurements are labeled by $y\in\mathcal{Y}$, and his outcomes $b\in\mathcal{B}$. Device-independent statements only rely on the family of probability distributions (the ``\textit{behavior}'') $\mathcal{P}=\{P(a,b|x,y)|a\in\mathcal{A},b\in\mathcal{B},x\in\mathcal{X},y\in\mathcal{Y}\}$.

As the title of the original paper goes, the family of inequalities derived by Collins, Gisin, Linden, Massar and Popescu (CGLMP) was meant to detect the nonlocality of high-dimensional quantum systems \cite{cglmp}. These inequalities, that have two inputs and $d$ outputs for both parties ($x,y\in\{0,1\}$,  $a,b\in\{0,1,2,\ldots,d-1\}$), are therefore natural candidates for dimension witnessing: indeed, the first example of a DIDW was based on CGLMP$_3$ \cite{brunner}, and semi-device independent witnessing of dimensions up to 20 was reported using the CGLMP family \cite{dada}. The DIDW character of CGLMP$_4$ was studied more recently: it was found that a violation greater than $I_4=0.315$ lower-bounds the dimension of the measured system to entangled ququarts~\cite{cy}.

One of the behaviors that exceeds the latter bound is $\mathcal{P}_{MES4}$ obtained by taking the maximally entangled state (MES) of two ququarts $|\Phi_4\rangle_{AB}=\frac{1}{2}(|00\rangle+|11\rangle+|22\rangle+|33\rangle)$ and performing the local projective measurement on the bases \cite{optmea,zohren}
\begin{align}
\label{eqn:optm}
\ket{a_x}&=\sum_{k=0}^{3}\frac{\exponent^{i\frac{\pi}{2}ak}}{2}\exponent^{ik\alpha_x}|k\rangle,\\
\label{eqn:optm2}
|b_y\rangle&=\sum_{k=0}^{3}\frac{\exponent^{-i\frac{\pi}{2}bk}}{2}\exponent^{ik\beta_y}|k\rangle,
\end{align}
with $\alpha_0=0$, $\alpha_1=\frac{\pi}{4}$, $ \beta_0=-\frac{\pi}{8}$, $\beta_1=\frac{\pi}{8}$. Indeed, one would find $I(\mathcal{P}_{MES4}) \approx 0.336$ \cite{cy}. 

Consider now the following encoding of a ququart into two qubits:
\begin{equation}
\label{eqn:encode}
\begin{split}
\ket{0} \mapsto \ket{0}\otimes\ket{0},\ & |1\rangle \mapsto \ket{0}\otimes\ket{1} \\
|2\rangle \mapsto \ket{1}\otimes\ket{0},\ & |3\rangle \mapsto \ket{1}\otimes\ket{1}\,.
\end{split}
\end{equation}
If both Alice and Bob perform this encoding, it is well known that the MES is mapped to the product of two two-qubit MESs: $|\Phi_4\rangle_{AB} \mapsto |\phi^+\rangle_{A_1B_1}\otimes|\phi^+\rangle_{A_2B_2}$, with $|\phi^+\rangle=\frac{1}{\sqrt{2}}(|00\rangle+|11\rangle)$. But here, also the optimal measurement bases \eqref{eqn:optm} and \eqref{eqn:optm2} factor as \textit{sequential measurements}. For Alice's, Bob's being analog, it reads:
\begin{align*}
\ket{0_x}_A&\mapsto\ket{+_{2\alpha_x}}_{A_1}\otimes\ket{+_{\alpha_x}}_{A_2}\\
\ket{1_x}_A&\mapsto\ket{-_{2\alpha_x}}_{A_1}\otimes\ket{+_{\alpha_x+\frac{\pi}{2}}}_{A_2}\\
\ket{2_x}_A&\mapsto\ket{+_{2\alpha_x}}_{A_1}\otimes\ket{-_{\alpha_x}}_{A_2}\\
\ket{3_x}_A&\mapsto\ket{-_{2\alpha_x}}_{A_1}\otimes\ket{-_{\alpha_x+\frac{\pi}{2}}}_{A_2}
\end{align*}
where $\ket{\pm_\varphi}=\frac{1}{\sqrt{2}}(\ket{0}\pm e^{i\varphi}\ket{1})$ are the eigenstates of $\sigma_{\varphi}=\cos\varphi\,\sigma_x+\sin\varphi\,\sigma_y$. Explicitly, this means that one can produce the behavior $\mathcal{P}_{MES4}$ by the following sequential strategy:
\begin{enumerate}
    \item The source sends out a pair of maximally entangled qubits. Given $x$, Alice measures her qubit in the basis $\ket{\pm_{2\alpha_x}}$; given $y$, Bob measures his qubit in the basis $\ket{\pm_{2\beta_y}}$.
    \item Later, the source sends out a second pair of qubits. If Alice obtained the outcome $+$ in her first measurement, she now measures the second qubit in the basis $\ket{\pm_{\alpha_x}}$; if she obtained $-$, in the basis $\ket{\pm_{\alpha_x+\frac{\pi}{2}}}$. Bob follows the analog procedure.
\end{enumerate}
In particular, one ends up certifying dimension four on both sides, where only a two-qubit source (admittedly used twice), single qubit local manipulations and classical feed-forward were implemented. For a sequential violation of the CGLMP inequality when the number of outcomes $d$ that is a power of two, see also~\cite{Lo2016}. In SM I, we show that the qutrit dimension witness based on CGLMP$_3$~\cite{brunner} may also be violated using solely two-qubit sources and sequential single-qubit measurements.

Hoping to better capture the experimental effort in proper high dimensional quantum experiments rather than sequential procedures, we introduce the notion of \emph{dimension irreducible by sequential operations}, or simply \emph{irreducible dimension}, of a quantum behaviour; and we provide an example of a DIDW that certifies irreducible dimension four.

\section{Correlations from sequential d-dimensional systems}
\label{section:CEM}

Let us define a sequential $d$-dimensional model as consisting of:
\begin{itemize}
\item $d$-dimensional sources (dimensionality),
\item operations and measurements performed sequentially on each $d$-dimensional system, possibly feeding forward the measurement outcomes (sequentiality), and
\item arbitrary local classical processing and shared randomness.
\end{itemize}

We then say that a behavior is \textit{sequential d-dimensional compatible} if it can be obtained with each involved party individually following a sequential $d$-dimensional model. The set of sequential $d$-dimensional compatible behaviors can be seen to be the closure-under-wiring~\cite{Allcock09,Lang14} of the set of $d$-dimensional quantum correlations~\cite{Navascues15a,Navascues15b}. Unfortunately, few explicit sets are known to be closed under wiring, and in general it is not known how to characterize the closure-under-wiring of a given set~\cite{Allcock09,Lang14}.

The smallest $d$ such that the behavior is sequentially $d$-compatible, is called the dimension irreducible by sequential operations, or simply the \emph{irreducible dimension} of the behavior. Every behavior that can be simulated with classical resources, including those describing prepare-and-measure schemes, has irreducible dimension 1. Just as for entanglement or nonlocality, nontrivial irreducible dimension must necessarily involve more than one party. The behavior $\mathcal{P}_{MES4}$ has irreducible dimension 2.

Given the previous example, one may fear that any probability distribution can be achieved by combining sufficiently many sequential measurements on qubit systems, rendering irreducible dimension witnessing a somewhat trivial exercise also in the quantum case. Fortunately, this is not the case: there exist distributions that lower bound the dimension of the involved devices to more than two even when sequential strategies are considered.

\section{A quantum behavior with irreducible dimension four}
\label{section:BSM}

A witness for irreducible dimension four must rule out sequential measurements on consecutive qubits and qutrits. We do not know how to express all of these constraints as a function of the observed probability distributions $P(a,b|x,y)$ in simple terms.  
In order to construct an example, we notice that an \textit{entangled measurement}, one whose eigenvectors are entangled states, cannot be sequential, since an entangled measurement cannot be achieved even with bidirectional classical communication. Besides, the minimal dimension to have entanglement is $d=4$ (two qubits). Thus, certification of such a measurement guarantees that a four-dimensional non-sequential operation is performed.

The possibility of certifying entangled measurements was demonstrated in the entanglement-swapping configuration, i.e.~in a tripartite scenario, either assuming knowledge of the dimensions \cite{PhysRevA.83.062112} or in the fully device-independent setting \cite{rabelo}. Exploiting a recent result on self-testing \cite{xy}, we construct an explicit behavior for a bipartite scenario, such that one of Bob's measurements can be certified to be entangled. 

Our behavior, denoted $\mathcal{P}_{BSM}=\{P(a,b|x,y): a,b,x \in \{0,1,2,3\}\,;\, y\in\{0,1,2,3,4\}\}$, uses four measurements for Alice and five measurements for Bob, each having four possible outcomes. The entangled measurement will be $y=4$. We need the other measurements to first establish that both Alice's and Bob's systems are composed of two subsystems in a local separable state, which cannot be assumed \textit{a priori} in a device-independent setting.

Let us first leave $y=4$ aside. The parties label each of the four-valued input and outcomes $a,b,x,y\in\{0,1,2,3\}$ as two bits: $c=2c_1+c_2\rightarrow (c_1,c_2)$. If
\begin{equation}\label{pfour}
P(a,b|x,y)=\prod_{i=1,2}P_{2\sqrt{2}}(a_i,b_i|x_i,y_i)\;,\;x,y\in\{0,1,2,3\}
\end{equation} where $P_{2\sqrt{2}}$ is the unique probability point that violates maximally the CHSH inequality, then the state shared between Alice and Bob is self-tested to the product $|\phi^+\rangle_{A_1B_1}\otimes|\phi^+\rangle_{A_2B_2}$ of two maximally-entangled two-qubits states \cite{xy}. Thus \eqref{pfour} certifies that there are indeed two subsystems in a separable state, both on Alice's side (denoted $A_1$ and $A_2$) and on Bob's ($B_1$ and $B_2$). In this situation, if $A_1$ and $A_2$ are found entangled conditioned on the outcome of $y=4$, then $y=4$ must be an entangled measurement on $B_1$ and $B_2$.

In order to test entanglement on Alice's side, we need suitable measurements, local on her subsystems --- and we have got them already. Indeed, it is an important feature of self-testing that not only the state, but also the measurements on the subsystems are self-tested as the optimal measurements for $|\phi^+\rangle$ to violate the CHSH inequality \cite{xy}. Thus we know that Alice's measurements are $\sigma_z\otimes \sigma_z$, $\sigma_x\otimes \sigma_z$, $\sigma_z\otimes \sigma_x$ and $\sigma_x\otimes \sigma_x$, up to local isometries.

Now we have a simple recipe to finish the construction of $\mathcal{P}_{BSM}$: for the measurement labelled $y=4$ we choose the Bell-State Measurement (BSM) on $B_1$ and $B_2$, which prepares $A_1$ and $A_2$ in states that violate CHSH maximally for those measurements of Alice. Thus we'll have
\begin{equation}\label{pfive}
P(a,b|x,4)=\frac{1}{4}P_{2\sqrt{2},b}(a_1,a_2|x_1,x_2)
\end{equation} since, as explained in the Supplemental Material, one must use a different CHSH expression for each value of Bob's outcome $b$. All the details are given in SM II. 

In summary, if one observes ${\cal P}_{BSM}$ defined by \eqref{pfour} and \eqref{pfive}, then the density matrix and measurement operators are acting locally on $\mathbb{C}^d$ with $d \geq 4$, and the statistics cannot be reproduced by Alice and/or Bob sequentially measuring several smaller-dimensional (qubit or qutrit) sources in their respective labs. In short, the behavior ${\cal P}_{BSM}$ has irreducible dimension greater than or equal to four.

\section{A witness of irreducible dimension four}\label{sec:witness}

In the previous section, ${\cal P}_{BSM}$ is just one behavior, i.e.~a single point in probability space: as such, it will never be observed exactly. In order to have a robust witness of irreducible dimension four, we need to demonstrate that when the observed probability point is not exactly ${\cal P}_{BSM}$, one party is still performing an entangled measurement.

It is clear that there is a large room for robustness in the $P(a,b|x,4)$, i.e.~in the choice of the entangled measurement itself: any behavior that shows a violation of CHSH (not necessarily maximal) for at least one value of $b$ would do. It is less easy to relax the self-testing part \eqref{pfour}, because one immediately loses the sharp conclusion on the existence of subsystems. In the absence of well-defined subsystems, the notion of entangled measurement becomes blurred.

In order to estimate the robustness of the criterion, we presume the existence of subsystems $B_1$ and $B_2$ on Bob's side and assume that the projectors of the first four measurements of Bob are of the form:
\begin{equation}
    \label{eq:pdtpjt}
    \Pi^{B}_{b|y}=\Pi^{B_1}_{b_1|y_1}\otimes\Pi^{B_2}_{b_2|y_2}
\end{equation}
where $\{\Pi^{B_1}_{b_1|y_1}\}_{b_1,y_1=0,1}$ is a two outcome projective measurement on $B_1$, $\{\Pi^{B_2}_{b_2|y_2}\}_{b_2,y_2=0,1}$ is a two outcome projective measurement on $B_2$, and with $c=2c_1+c_2$ as before. Physically, this is equivalent to Bob measuring $B_1$ and $B_2$ independently and concatenating the two outcomes into one outcome string.

Under this assumption, we now demonstrate the robustness of the certification of entangled measurements and hence of irreducible dimension. For this, we make use of the SWAP technique~\cite{Yang14,Bancal15}. A so-called SWAP operator, defined in terms of the parties' measurements, is used to exchange a particular part of the measured system with an auxiliary system of trusted dimension. Any linear function of the resulting quantum state can then be bounded over all quantum realizations through the NPA hierarchy of semi-definite programming (SDP)~\cite{PhysRevLett.98.010401, 1367-2630-10-7-073013}.



Here, we thus consider two external qubits registers for Bob together with two qubit SWAP operators~\cite{Yang14}, each operator swapping subsystem B1, respectively B2, with one of the external qubit registers. The resulting double-SWAP operator on Bob's system can be expressed in terms of Bob's measurement operators $\Pi^{B}_{b|y}$ as
\begin{equation}\label{eq:bigswap}
\mathcal{S}_{BB'}\ket{i,j}_{B'} = \sum_{k,l=0}^1 \ket{k,l}_{B'} X_{k,l} \Pi^{B}_{f(i,j,k,l)|0} X_{i,j}
\end{equation}
where $f(i,j,k,l) = 2\,(i\oplus k)+ (j\oplus l)$, $X_{i,j} = \sum_{k=0}^3 (-1)^{jk+i\lfloor k/2\rfloor}\Pi^{B}_{k|3}$ and $\oplus$ is the sum modulo 2 (c.f. SM III). 

We then estimate whether Bob's fifth measurement is entangled by computing
\begin{equation}\label{eq:obj}
F = \frac{1}{4}\sum_{i=0}^3 \Tr\left[\Pi_{i|4}^B \mathcal{S}_{BB'} (\rho_{AB}\otimes\ket{\varphi_i}_{B'}\bra{\varphi_i})\mathcal{S}_{BB'}^\dag\right],
\end{equation}

where $\ket{\varphi_i}=\sum_j(-1)^{ij}\ket{j}\otimes\ket{\lfloor i/2\rfloor\oplus j}$ are the four Bell states. This expression can be understood as follows: the SWAP operator places a maximally entangled state in $B_1\otimes B_2$, after which one checks how close $\Pi^{B}_{i|4}$ is to $\ket{\varphi_i}\bra{\varphi_i}$. In SM III we show that this fidelity is related to the fidelity of an entanglement swapping protocol that used Bob's last measurement to entangle two remote qubits; when $F>1/2$ at least one of Bob's measurement operators must be entangled. This conclusion is contingent on the assumption \eqref{eq:pdtpjt} made on Bob's system, which ensures that the SWAP operator~\eqref{eq:bigswap} factorises according to $\mathcal{S}=\mathcal{S}_I\otimes\mathcal{S}_{II}$ where $\mathcal{S}_{I}$ acts on $B_1$ and the auxiliary system $B_1'$ and $\mathcal{S}_{II}$ acts on $B_2$ and $B_2'$. An example showing the importance of this assumption for the presented argument is given in SM III. 

\begin{figure}[t]
    \centering
    \includegraphics[width=0.5\textwidth]{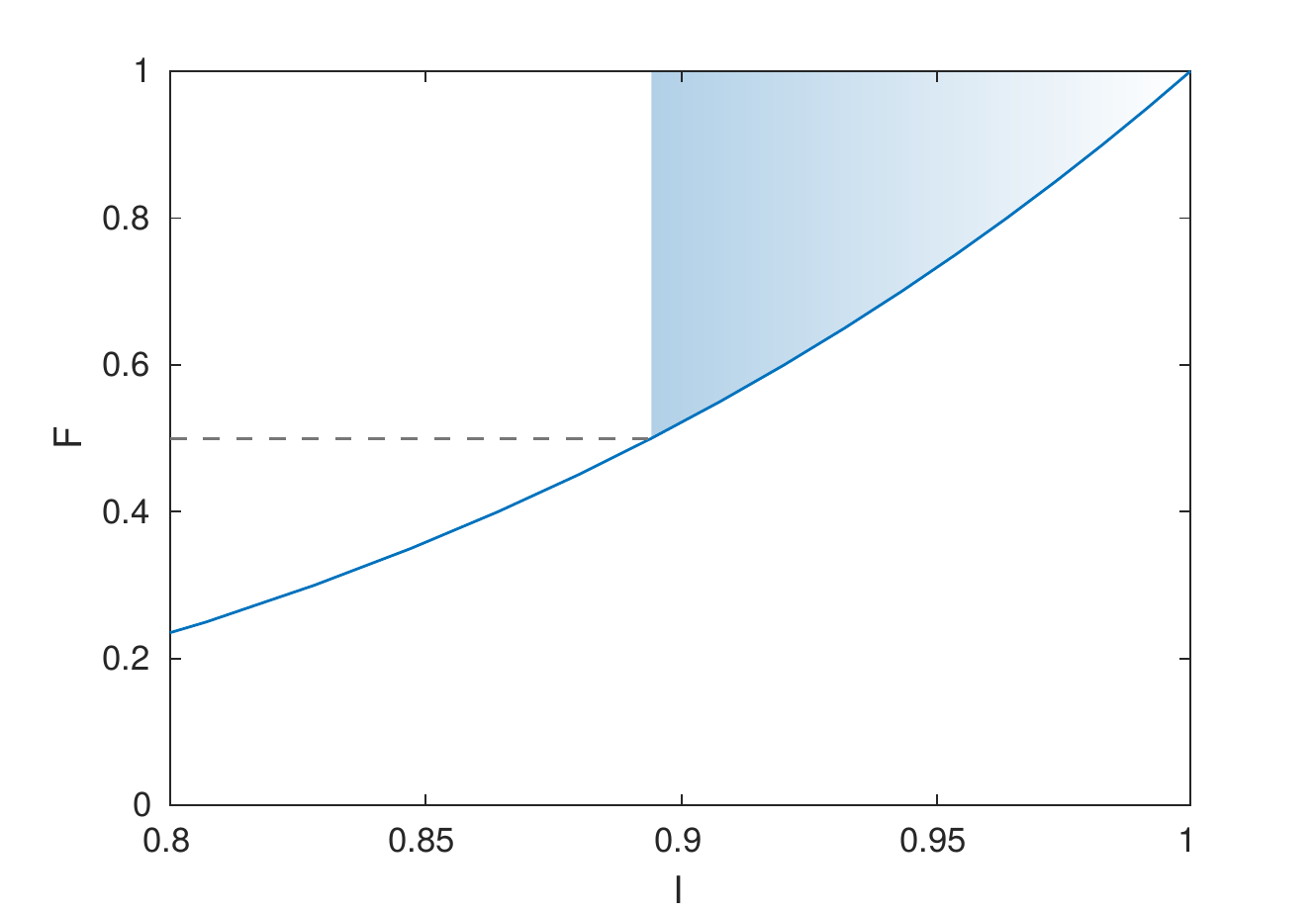}
    \caption{Lower bound on the fidelity $F$ as a function of the violation of the Bell inequality $I\leq 0$. The shaded area highlights values of $I$ above $I^*\simeq 0.8942$, which guarantee that $F>1/2$. The maximal quantum violation of $I$ is $I=1$.}
    \label{fig:InequalityBound}
\end{figure}

To bound the quantity $F$ over all possible quantum realizations which are compatible with some behavior $\mathcal{P}$, we constructed an SDP matrix of size $390\times390$ corresponding to a relaxation of the Navascu{\'e}s-Pironio-Ac{\'i}n (NPA) hierarchy \cite{PhysRevLett.98.010401,*1367-2630-10-7-073013}. We then minimized $F$ over all such matrices which are compatible with the chosen quantum behavior. For the sake of an example, let us consider the behavior obtained with a perfect implementation of the measurements on the tensor product of two two-qubit Werner states $\left(V\ket{\phi^+}\bra{\phi^+} + (1-V)\mathbb{I}/4\right)^{\otimes 2}$, resulting in a noisy version of $\mathcal{P}_{BSM}$. We find $F>1/2$ for $V\gtrsim 0.987$ (c.f. SM IV). The corresponding dual SDP program provides a certificate for this conclusion in the form of a bipartite Bell inequality.
This conclusion is readily confirmed by computing the minimal fidelity $F$ which is compatible with some violation of the inequality $I$ (c.f.~Figure \ref{fig:InequalityBound}).

\section{Conclusion}

In this paper, we pointed out that some device-independent dimension witnesses can be violated with lower dimensional systems and sequential measurements on them. This somehow defeats the operational goal of these witnesses, which is not simply to prove that some dimensionality ``is there'', but rather to certify that one can do quantum information processing. The same concern should be raised also for non-DI dimension witnesses: for instance, the lower bounds of Ref.~\cite{sikora} are multiplicative for product correlations, so by just using a qubit source $n$ times they certify dimension $2^n$.

Then we showed that this obstacle can be overcome: it is possible to construct witnesses that capture a more appropriate notion of dimension, namely what we called the irreducible dimension of a quantum behavior. This solution was based on an example of entangled measurements certification. From now onwards, in the presence of a dimension witness, it will be important to check which irreducible dimension it certifies. 

Some problems remain open. The robustness of our criterion was proved under some additional assumptions, because we have not found a way of identifying subsytems in a device-independent setting. Alternatively, one may think of approaches that are based on different criteria. It would also be interesting to investigate the case where the sequentiality assumption that we used here is removed. Ruling out that low-dimensional states and operations can be responsible for some observed behavior, independently of the way in which these resources are combined would then lead to witnessing behaviors with genuine dimension $d$. A similar problem in the context of entanglement theory with characterized devices was recently considered by Kraft et al.~\cite{kraft}. In the ideal case, this work can be made device-independent by using self-testing~\cite{CGS}. Independently of these questions, it would also be interesting to obtain a compact characterization of the statistics achievable with sequential measurements. This might provide an alternative approach to study the closure-under-wiring set of correlations.


\section*{Acknowledgments} We thank Nicolas Brunner, Jiangbin Gong, Otfried Guehne and Miguel Navascu\'es for feedback and discussions. This research is supported by the Singapore Ministry of Education Academic Research Fund Tier 3 (Grant No. MOE2012-T3-1-009); by the National Research Fund and the Ministry of Education, Singapore, under the Research Centres of Excellence programme; by the Swiss National Science Foundation (SNSF), through the NCCR QSIT and the Grant number PP00P2-150579.

\pagebreak
\onecolumngrid
\appendix
\section{Violation of qutrit dimension witness based on CGLMP$_3$ inequality}\label{appcglmp3}

We will show how the qutrit dimension witness proposed in \cite{brunner} can be violated using three pairs of maximally entangled qubits and sequential qubit measurements. We do so by first generating the statistics that violates the CGLMP$_8$ inequality, then locally coarse grain to three outcomes to test for the CGLMP$_3$ violation. 

Using the standard binary encoding, the maximally entangled qu-8it can be factorised into 3 pairs of maximally entangled qubits:
\begin{multline*}
(|00\rangle+|11\rangle+|22\rangle+|33\rangle+|44\rangle+|55\rangle+|66\rangle+|77\rangle)_{A,B} \\ \mapsto (|00\rangle+|11\rangle)_{A_1,B_1}\otimes(|00\rangle+|11\rangle)_{A_2,B_2}\otimes(|00\rangle+|11\rangle)_{A_3,B_3}.
\end{multline*}
The measurement bases of Alice and Bob will also factorise such that they can be done sequentially on the three qubits. Following the notation in the main text, the behavior $\mathcal{P}_{MES8}$ can be produced by the following sequential strategy by Alice (Bob being the analog): the source sends out three pairs of maximally entangled qubits sequentially. Alice measures her first qubit in the basis $\ket{\pm_{4\alpha_x+\pi}}$. If Alice obtained the outcome $+$ in her first measurement, she measures the second qubit in the basis $\ket{\pm_{2\alpha_x}}$; if she obtained $-$, in the basis $\ket{\pm_{2\alpha_x+\frac{\pi}{2}}}$; on her third qubit, Alice measures in the basis $\ket{\pm_{\alpha_x}}$, $\ket{\pm_{\alpha_x+\frac{\pi}{2}}}$, $\ket{\pm_{\alpha_x+\frac{\pi}{4}}}$ or $\ket{\pm_{\alpha_x+\frac{3\pi}{4}}}$, if the outcomes of her previous two measurements were $``+,+"$, $``+,-"$, $``-,+"$ or $``-,-"$ respectively. 

\begin{figure}[ht]
    \centering
    \includegraphics[width=0.35\textwidth]{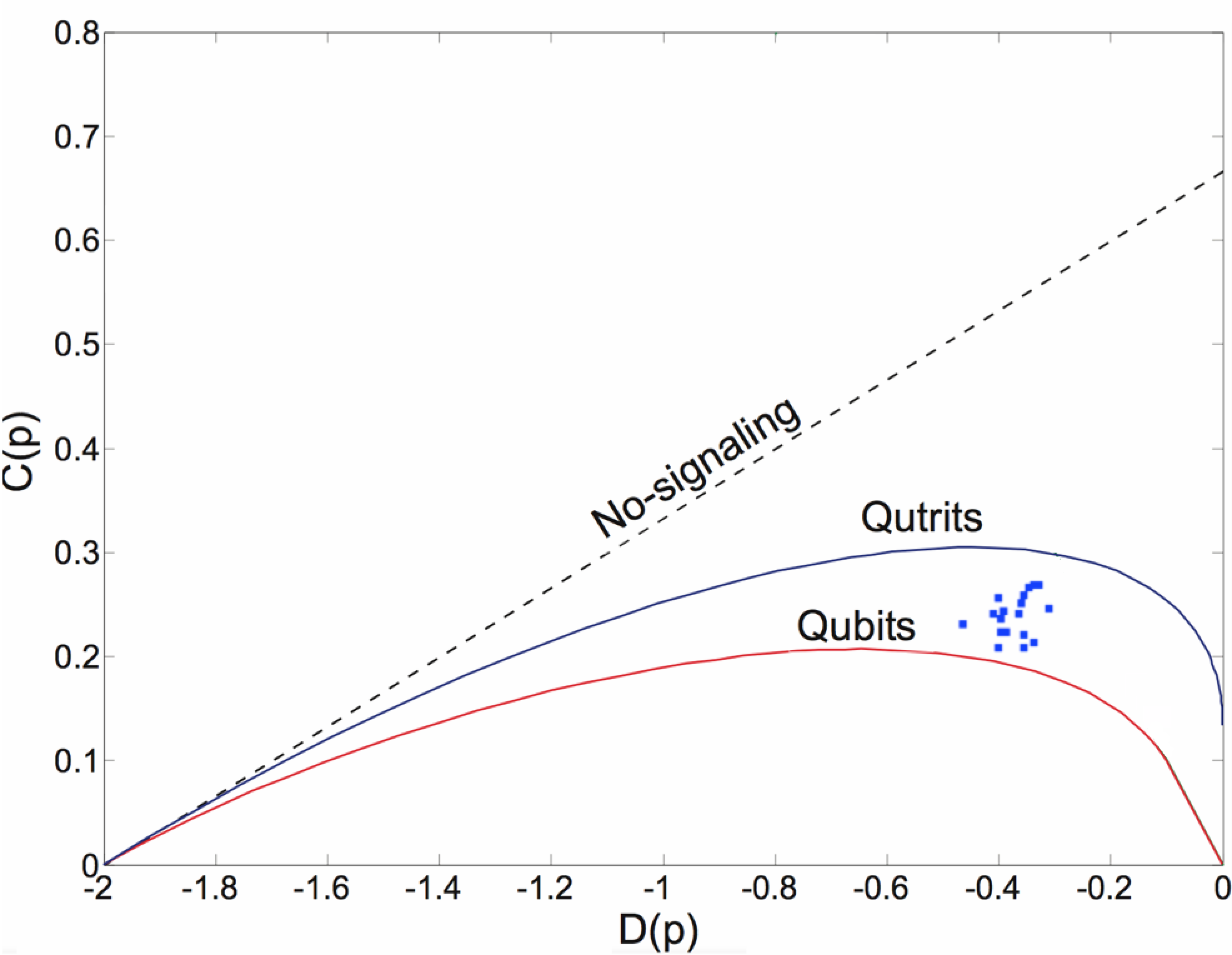}
    \caption{A slice of the 2-input, 3-outcome no signalling polytope. The vertical axis of this figure \cite{brunner} is the amount of violation of the CGLMP$_3$ inequality. The red curve is a qutrit DIDW: points above the curve can only be achieved using entangled quantum systems of dimension at least $3$. Data points in blue are those that can be achieved using three pairs of entangled qubits.}
    \label{fig:cglmp3}
\end{figure}

The resultant eight outcome distribution can be coarse grained into a three outcome distribution by grouping the outcomes. It can be verified that under the relabeling
\begin{align*}
{1,4,7}\mapsto{0} \\ {2,5}\mapsto{1} \\ {0,3,6}\mapsto{2},
\end{align*}
one can achieve a violation of $0.2677$ for the CGLMP-3 inequality. This violates the qutrit dimension witness based on the CGLMP-3 inequality \cite{brunner}, $I_{qubit} \leq (\sqrt(2)-1)/2 \approx 0.2071$. We have also found more ad hoc points which violate the dimension witness, shown in Fig.~\ref{fig:cglmp3}.

\section{Details on $\mathcal{P}_{BSM}$}\label{appbsm}


In this appendix, we show the state and measurements to obtain $\mathcal{P}_{BSM}$. We may assume that Alice and Bob share two pairs of maximally entangled qubits in the form $|\psi\rangle_{AB}=|\phi^+\rangle_{A_1B_1}\otimes|\phi^+\rangle_{A_2B_2}$. For the first four measurements $x,y \in \{0,1,2,3\}$, Alice and Bob interpret the four inputs as two binary inputs for the two subsystems: $x = 2x_1+x_2, y = 2y_1+y_2$. The measurements they perform are as follows (for $x_i,y_i \in \{0,1\}$):
\begin{align*}
&x_1: \sigma_z, \sigma_x,  && y_1:\frac{\sigma_z+\sigma_x}{\sqrt{2}}, \frac{\sigma_z-\sigma_x}{\sqrt{2}}, \\
&x_2: \frac{\sigma_z+\sigma_x}{\sqrt{2}}, \frac{\sigma_z-\sigma_x}{\sqrt{2}},  && y_2: \sigma_z, \sigma_x,
\end{align*}
the measurements corresponding to setting $x_i$ ($y_i$) being performed on system $A_i$ ($B_i$).

The two binary outcomes combine to form a quarternary outcome via $a = 2a_1+a_2, b = 2b_1+b_2$. Overall, $P(a,b|x,y)$ is given by Eq. (4) with
\begin{equation}\label{probchsh}
P_{2\sqrt{2}}(a_i,b_i|x_i,y_i)=\frac{1}{4}\left(1+(-1)^{f_0}\frac{1}{\sqrt{2}}\right)
\end{equation}
with $f_0=a_i\oplus b_i-x_iy_i$ and $\oplus$ represents sum modulo 2.

The last measurement of Bob, $y=4$, is a Bell state measurement, corresponding to projection on the following basis:
\begin{equation}
\begin{split}
\label{eqn:bsm}
b=0: |\phi^+\rangle=\frac{1}{\sqrt{2}}(|00\rangle+|11\rangle),\\
b=1: |\phi^-\rangle=\frac{1}{\sqrt{2}}(|00\rangle-|11\rangle),\\
b=2: |\psi^+\rangle=\frac{1}{\sqrt{2}}(|01\rangle+|10\rangle),\\
b=3: |\psi^-\rangle=\frac{1}{\sqrt{2}}(|01\rangle-|10\rangle).
\end{split}
\end{equation}
With this choice, conditioned on the outcome $b$, the Bell expression $S_b$ that takes the value $2\sqrt{2}$ is:
\begin{equation}
\begin{split}
S_0=-S_3=E_{00}+E_{01}+E_{10}-E_{11}\,.\\
S_1=-S_2=E_{00}+E_{01}-E_{10}+E_{11}\,.
\end{split}
\end{equation}
Thus we obtain Eq.(5) where the $P_{2\sqrt{2},b}$ have the same form as \eqref{probchsh}, with $f_3=f_0\oplus 1$, $f_1=a_i\oplus b_i-(x_i\oplus 1)y_i$ and $f_2=f_1\oplus 1$.

\section{Entangled measurement and entanglement swapping}\label{appfid}
Here, we relate the figure of merit used in section V of the main text to the singlet fidelity of the state that can be created by using Bob's fifth measurement $B_4$ in an entanglement swapping experiment involving two singlet states. Since this fidelity can only be high if the applied measurement is entangled, this justifies the usage of the fidelity of the main text to detect the entangled character of Bob's measurement.

\begin{figure}[t]
    \centering
    \includegraphics[width=0.3\textwidth]{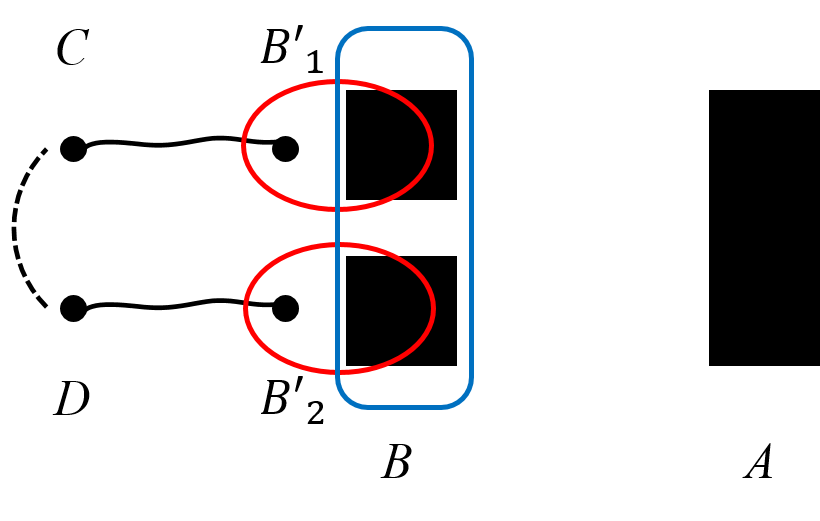}
    \caption{Schematic of the thought experiment. Besides the measured system $A$ and $B$, Bob also has access to two pairs of maximally entangled qubits $\ket{\phi^+}_{B_1'C}\otimes\ket{\phi^+}_{B_2'D}$. After the SWAP operator $S_{BB'}$ is applied (the red circles), the singlet fidelity of $\rho_{CD}$ conditioned on the outcome of Bob's fifth measurement (the blue rectangle) is denoted as $F_i$. The average fidelity $F = \sum_i P_iF_i$ is bounded through the NPA hierarchy.}
    \label{fig:entg_swapping}
\end{figure}

We consider a thought experiment involving our two parties together with two auxiliary maximally entangled two-qubit states $\ket{\phi^+}_{B_1'C}\otimes\ket{\phi^+}_{B_2'D}$, see Fig.~\ref{fig:entg_swapping}. Giving the systems $B'=(B_1',B_2')$ to Bob, we allow him to perform arbitrary operations between these two qubits and his system. He is not allowed to access systems $C$ and $D$ however. The idea is then to ask whether there is an operation that Bob can do on $(B,B')$, which necessarily results in the two qubits $C$ and $D$ being entangled. If this is the case, then he can perform entanglement swapping and this shows that Bob is able to perform an entangled measurement, i.e. a 4-dimensional non-sequential operation.

Concretely, we would like to let Bob apply his fifth measurement on the two auxiliary systems $B_1'$ and $B_2'$. In order to allow for this, Bob first needs to swap the two auxiliary qubits with part of his internal system $B$ (on which $B_4$ has a non-trivial action)~\cite{Yang14}. As mentioned in the main text, the assumption (6) in the main text allows us to apply two independent qubit swaps $\mathcal{S}_I$ and $\mathcal{S}_{II}$ here, which swaps $B_1'$ with part of $B_1$ and $B_2'$ with part of $B_2$ respectively. To see this, we write the SWAP operator in Eq. (7) explicitly as $\mathcal{S} = (U_IV_IU_I)(U_{II}V_{II}U_{II})$ where
\begin{align*}
U_I&=\ket{0}_{B_1'}\bra{0}+(\Pi_{0|3}+\Pi_{1|3}-\Pi_{2|3}-\Pi_{3|3})\ket{0}_{B_1'}\bra{0},
&U_{II}&=\ket{0}_{B_2'}\bra{0}+(\Pi_{0|3}-\Pi_{1|3}+\Pi_{2|3}-\Pi_{3|3})\ket{0}_{B_2'}\bra{0}, \\
&=[~\mathbb{I}_{B_1}\otimes\ket{0}_{B_1'}\bra{0}+(\Pi^{B_1}_{0|1}-\Pi^{B_1}_{1|1})\otimes\ket{1}_{B_1'}\bra{1}~]_I\otimes\mathbb{I}_{II},
& & =\mathbb{I}_I\otimes[~\mathbb{I}_{B_2}\otimes\ket{0}_{B_2'}\bra{0}+(\Pi^{B_2}_{0|1}-\Pi^{B_2}_{1|1})\otimes\ket{1}_{B_2'}\bra{1}~]_{II}\\
V_I & =\Pi_{0|0}+ \Pi_{1|0}+(\Pi_{2|0}+\Pi_{3|0})~\sigma_{x,B_1},
&V_{II}&=\Pi_{0|0}+ \Pi_{2|0}+(\Pi_{1|0}+\Pi_{3|0})~\sigma_{x,B_2},\\
& = [~\Pi^{b_1}_{0|0}\otimes\mathbb{I}_{B_1'}+\Pi^{b_1}_{1|0}\otimes\sigma_{x,B_1'}~]_I\otimes\mathbb{I}_{II}, 
&&=\mathbb{I}_I\otimes[~\Pi^{b_2}_{0|0}\otimes\mathbb{I}_{B_2'}+\Pi^{b_2}_{1|0}\otimes\sigma_{x,B_2'}~]_{II}.
\end{align*}
Hence $U_I$ and $V_I$ can be viewed as operators on subsystem I and $U_{II}$ and $V_{II}$ as operators on subsystem II. The operator $\mathcal{S}$ thus factorises as promised. 

To highlight the importance of this assumption, we note that in a situation in which the assumption is not met, the particular SWAP operator chosen here may lead to entanglement swapping when Bob's fourth measurement is separable across $B_1$ and $B_2$. A simple example is the case when the measurements $y=0$ and $y=3$ are respectively $\sigma_z\otimes\sigma_z$ and $\sigma_x\otimes\sigma_x$, with the encodings $(0,0)\rightarrow 0, (0,1)\rightarrow 3, (1,0)\rightarrow 2, (1,1)\rightarrow 1$, and $(0,0)\rightarrow 0, (0,1)\rightarrow 1, (1,0)\rightarrow 3, (1,1)\rightarrow 2$, for the outcomes. Using Eq. (7) as the SWAP operator will result in a fidelity $F=1$ in Eq. (8) when the measurement $y=4$ is $\sigma_z\otimes\sigma_x$, which is not entangling.

Bob's measurement $B_4$ has four possible outcomes. The singlet fidelity of the state produced by the measurement operator corresponding to each outcome can be computed by using the appropriate reference state $\ket{\varphi_0}=(\ket{00}+\ket{11})/\sqrt{2}$, $\ket{\varphi_1}=(\ket{00}-\ket{11})/\sqrt{2}$, $\ket{\varphi_2}=(\ket{01}+\ket{10})/\sqrt{2}$ or $\ket{\varphi_3}=(\ket{01}-\ket{10})/\sqrt{2}$:
\begin{equation}
\overline F_i=\bra{\varphi_i} \Tr_{ABB'}\left[\Pi_{i|4}^B\mathcal{S}_{BB'} (\rho_{AB} \otimes \sigma_{B'CD})\mathcal{S}_{BB'}^\dag \right]\ket{\varphi_i},
\end{equation}
where $\sigma_{B'CD}=\ket{\psi^+}_{B_1'C}\bra{\psi^+}\otimes\ket{\psi^+}_{B_2'D}\bra{\psi^+}$ is the initial state of the auxiliary systems. These fidelities include the probability that outcome $i$ is produced:
\begin{equation}
\overline F_i = P_i F_i
\end{equation}
where $P_i=\Tr\left[\Pi_{i|4}^B\mathcal{S}_{BB'} (\rho_{AB} \otimes \sigma_{B'CD})\mathcal{S}_{BB'}^\dag \right]$ is the probability of observing outcome $i$ in this thought experiment and $F_i$ is the singlet fidelity of the (normalized) state produced by the measurement operator $\Pi_{i|4}^B$.

Since $\sum_i P_i=1$, the average singlet fidelity after entanglement swapping is then
\begin{equation}
F = \sum_{i=0}^3 P_i F_i = \sum_{i=0}^3 \overline F_i.
\end{equation}
This fidelity can only be larger than $1/2$ if $F_i>1/2$ for at least one $i$. When this is the case, the measurement $\Pi_{i|4}^B$ is necessarily entangled.

It is useful to note that the above expressions can be reduced through the relation $\overline F_i=\tilde F_i/4$ with
\begin{equation}
\tilde F_i = \Tr\left[\Pi_{i|4}^B \mathcal{S}_{BB'} (\rho_{AB}\otimes\ket{\varphi_i}_{B'}\bra{\varphi_i})\mathcal{S}_{BB'}^\dag\right].
\end{equation}
Here, the fidelity $\tilde F_i$ can be interpreted as the overlap evaluated through the swap operator $\mathcal{S}_{BB'}$ between the tested measurement $\Pi_{i|4}$ and a single maximally entangled state of two qubits. We then have $F = \frac{1}{4}\sum_i \tilde F_i$, which is the figure of merit used in the main text.

Note that previous approaches certifying an entangled measurement relied on the tripartite entanglement swapping scenario~\cite{rabelo}. Here, we certify the entangled nature of a measurement in a bipartite scenario.

\section{Numerical study}\label{appsdp}

As discussed in the main text, we consider the following semi-definite program:
\begin{equation}
\begin{split}
\underset{\{x^i\}}{\text{min}}\ \ & \sum_i f_i x^i\\
\text{s.t.}\ \ & \Gamma \geq 0 \\
& \sum_i h_i(a,b|x,y) x^i = P(a,b|x,y)
\end{split}
\end{equation}
where $\Gamma = \sum_i G_i x^i$ is the NPA matrix~\cite{PhysRevLett.98.010401,*1367-2630-10-7-073013} corresponding to a certain hierarchy level, $h_i(a,b|x,y)$ are constants, $P(a,b|x,y)$ is the behavior under consideration, and $f_i$ are the coefficients defining the objective function (8) in terms of the moments $x^i$.

\begin{figure}[t]
    \centering
    \includegraphics[width=0.5\textwidth]{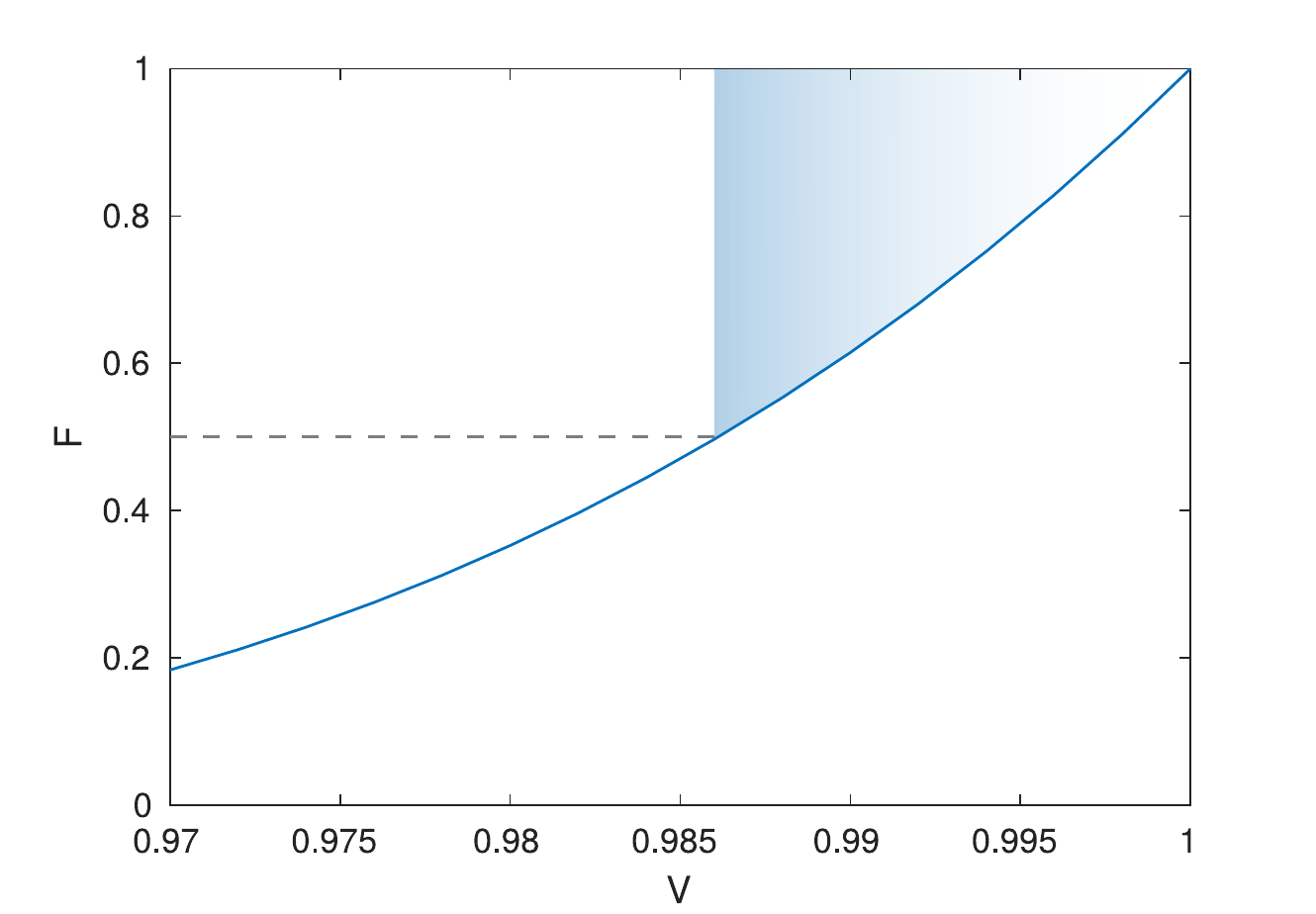}
    \caption{Lower bound on $F$ as a function of the Werner state fidelity. The shaded area highlights the range of $V$ for which a fidelity larger than $1/2$ can be certified. The inequality corresponding to $V=0.987$ is studied in the main text.}
    \label{fig:FidelityWerner}
\end{figure}

The behavior $P(a,b|x,y)$ that we consider here are the ones obtained upon measuring two Werner states with visibility $V$, i.e. $\rho = \left(V\ket{\phi^+}\bra{\phi^+} + (1-V)\mathbb{I}/4\right)^{\otimes 2}$ with Alice's four measurements and Bob's five measurement that give rise to $\mathcal{P}_{BSM}$ when $V=1$. In Figure~\ref{fig:FidelityWerner}, we show the result of this computation, i.e. the lower bound on $F$ that can be certified from the quantum statistics as a function of the single-pair visibility $V$. The moment matrix that we used here is generated by the following 390 operators: $\{\{\mathbb{I}, \Pi_{a'|x}^A\} \otimes \{\mathbb{I}, \Pi_{b'|y}^B, \Pi^B_{0|1}\Pi^B_{0|4},
\Pi^B_{1|4}\Pi^B_{1|1},
\Pi^B_{2|5}\Pi^B_{1|4},
\Pi^B_{0|4}\Pi^B_{2|1}\Pi^B_{2|4},
\Pi^B_{2|4}\Pi^B_{1|1}\Pi^B_{1|4},
\Pi^B_{0|5}\Pi^B_{1|1}\Pi^B_{0|4},
\Pi^B_{1|5}\Pi^B_{2|1}\Pi^B_{1|4},
\Pi^B_{1|5}\Pi^B_{0|4}\Pi^B_{2|1},
\Pi^B_{2|5}\Pi^B_{2|4}\Pi^B_{0|1},$
$\Pi^B_{0|5}\Pi^B_{1|4}\Pi^B_{0|1}\Pi^B_{1|4},
\Pi^B_{0|5}\Pi^B_{2|4}\Pi^B_{2|1}\Pi^B_{1|4},
\Pi^B_{1|5}\Pi^B_{1|4}\Pi^B_{0|1}\Pi^B_{2|4},
\Pi^B_{1|5}\Pi^B_{2|4}\Pi^B_{2|1}\Pi^B_{2|4},
\Pi^B_{2|5}\Pi^B_{1|4}\Pi^B_{1|1}\Pi^B_{0|4}
\}\}$, where $a',b'\in\{0,1,2\}$ and the kronecker product distributes over the sets, i.e. $\{x,y\}\otimes\{z,w\}=\{x\otimes z,x\otimes w,y\otimes z,y\otimes w\}$.

Even though this study shows some resistance to noise (it tolerated a visibility drop of close to 1\% for the considered settings), it remains sensitive to any change of statistics which cannot be attributed to a drop of visibility. In order to demonstrate that our conclusion is also robust to such changes, we extracted a Bell inequality from the SDP dual at $V=0.987$. Interestingly, the inequality only involves three of Bob's settings (corresponding to his measurements 0, 3 and 4). Here is the Collins-Gisin table representation of this inequality~\cite{gisinform}:\ \\

\begin{center}
\mbox{
$I=
\left(\begin{tabular}{c||ccc|ccc|ccc}
-3.2291 & -0.0541 & 0.2518 & 0.2518 & -11.6034 & -3.5924 & -3.5924 & 1.6345 & 0.0831 & 1.6588 \\
\hline\hline
-5.7609 & 6.4226 & 3.3537 & 3.3537 & 5.8055 & 2.8521 & 2.8521 & -0.1087 & -0.2512 & -0.0683 \\
-0.7976 & 3.2625 & 1.6663 & -1.1834 & 3.2034 & 1.9221 & -0.9452 & -1.7480 & -1.0715 & -0.8582 \\
-0.7976 & 3.2625 & -1.1834 & 1.6663 & 3.2034 & -0.9452 & 1.9221 & -1.7480 & -1.0715 & -0.8582 \\
\hline
-2.8990 & 0.0000 & 2.8490 & -2.8491 & 5.7978 & 2.8988 & 2.8988 & -0.0001 & -0.0001 & -0.0000 \\
-3.3050 & 3.2064 & 4.5539 & 1.7050 & 2.8989 & 1.7012 & -1.0498 & 0.0481 & 0.9004 & -0.8496 \\
-0.2630 & -3.2983 & 1.1714 & -1.6775 & 2.8989 & -1.2426 & 1.9254 & 0.1998 & 0.9062 & -0.8330 \\
\hline
-2.8990 & 0.0000 & -2.8491 & 2.8490 & 5.7978 & 2.8988 & 2.8988 & -0.0001 & -0.0001 & -0.0000 \\ 
-0.2630 & -3.2983 & -1.6775 & 1.1714 & 2.8989 & 1.9254 & -1.2426 & 0.1998 & 0.9062 & -0.8330 \\
-3.3050 & 3.2064 & 1.7050 & 4.5539 & 2.8989 & -1.0498 & 1.7012 & 0.0481 & 0.9004 & -0.8496 \\
\hline
-0.0463 & -6.4220 & -3.3526 & -3.3526 & 5.8056 & 2.9531 & 2.9531 & 0.1092 & 0.2516 & 0.0687 \\
2.3600 & -3.1590 & -4.5358 & -1.6862 & 2.6020 & 1.6719 & -1.1955 & -1.6392 & -0.8202 & -0.7897 \\
2.3600 & -3.1590 & -1.6862 & -4.5358 & 2.6020 & -1.1955 & 1.6719 & -1.6392 & -0.8202 & -0.7897 \\
\end{tabular}\right)\leq0$}
\end{center}\ \\

This inequality constitutes a certificate that the statistics with this visibility are only compatible with $F>1/2$. By its linearity, this Bell inequality also guarantees that any statistics achieving a larger or equal violation of the inequality must have $F > 1/2$ as well. This result is presented in Figure~1 of the main text. This plot was obtained by solving the following SDP:
\begin{equation}
\begin{split}
\underset{\{x^i\}}{\text{min}}\ \ & \sum_i f_i x^i\\
\text{s.t.}\ \ & \Gamma \geq 0 \\
& \sum_{abxyi} \alpha_{abxy} h_i(a,b|x,y) x^i = v
\end{split}
\end{equation}
where $I=\sum_{abxy}\alpha_{abxy}P(a,b|x,y)\leq 0$ is the considered Bell inequality and $v$ its observed value.

\bibliography{Bibliography}  
\end{document}